\documentclass[aps,twocolumn,floats,superscriptaddress,showpacs]{revtex4}

\usepackage{graphicx}

\begin{document}

\title{Understanding the Heavy Fermion Phenomenology from Microscopic
Model}

\author{Ping Sun}
\address{Department of Physics and Astronomy,
Rutgers University, Piscataway, NJ 08854-8019}

\author{Gabriel Kotliar}
\address{Department of Physics and Astronomy,
Rutgers University, Piscataway, NJ 08854-8019}

\date{\today}

\begin{abstract}
  We solve the 3D periodic Anderson model using a two impurity cluster
  DMFT. We obtain the temperature v.s. hybridization phase diagram.
  Approaching the quantum critical point (QCP) both the Neel and
  lattice Kondo temperatures decrease and they do not cross at the
  lowest temperature we reached. While strong ferromagnetic spin
  fluctuation on the Kondo side is observed, our result suggests the
  critical static spin susceptibility is local in space at the QCP. We
  observe in the crossover region logarithmic temperature dependence
  in the specific heat coefficient and spin susceptibility.
\end{abstract}


\maketitle

Heavy Fermion phenomenon is among the most intensively studied
subjects in condensed matter physics \cite{stewart}. Experimental
information accumulated over the past thirty years has revealed many
unconventional aspects of the heavy Fermion physics
\cite{stewart,coleman}.

Heavy Fermion physics is derived from a local moment band hybridizing
with an extended conduction band. The hybridization induces the
competing Kondo ($J_K$) and RKKY interactions \cite{doniach}. At $J_K
\ll T$, the physics is dominated by those of the two bands separately
with the hybridization as a perturbation. As the temperature is
lowered, depending on the strength of $J_K$, different physics may
develop. In the region where $J_K \ll W$ (the conduction bandwidth),
the RKKY interaction prevails. The RKKY is a long range exchange
mediated by the conduction electrons near the Fermi surface and
oscillates with $k=2k_F$ asymptotically. At $T \lesssim J_K^2/W$, the
RKKY interaction can induce a transition to a magnetically ordered
phase. In the crossover regime on this side, the Kondo behavior,
though subdominant, would still show up in various measurables due to
its non-analyticity in terms of the energy cutoff, e.g.
temperature. As $J_K$ is increased, the Kondo effect becomes more
important and eventually dominates. Here the Kondo screening begins at
high temperatures $T\sim J_K$ where the conduction electrons near the
Fermi energy starts to screen the local f-moments. If this remains so
as the temperature is lowered, there would not be enough conduction
electrons to completely screen the f-moment lattice
\cite{nozieres}. Actually in heavy Fermions the entire conduction
Fermi sea gets involved in screening. As the temperature is lowered,
conduction electrons farther away from the Fermi surface participate
in the Kondo screening. The system may then be described, to the
leading order, as a band of local moments whose magnitude is
progressively reduced. These reduced local moments still hybridize
with the conduction electrons which have not participated in the Kondo
screening and live near the conduction band bottom. Hence, on the
Kondo side, the RKKY correlation, as $T$ is lowered, becomes more
ferromagnetic (FM). The FM spin correlation remains at further lower
temperatures when the heavy Fermi liquid is formed. Macroscopically,
the FM behavior is related to the lattice Kondo energy, $T_0$, which
is proportional to the Fermi energy. $T_0$ can be defined in terms of
the saturated homogeneous spin susceptibility in the Fermi liquid
phase, $\chi_{\vec{k}=0}=C/T_0$, where $C$ is the Curie
constant. Should $T_0$ approach zero, strong FM spin fluctuation would
be observed.

It turns out that the thermodynamics related to the continuous
condensation of the local moments into the heavy fermion fluid is
quite universal \cite{pines}, as contrasted with that more material
specific in the low temperature region. At low temperatures, various
phases may develop, including a superconducting phase
\cite{stewart}. One interesting possibility is that the competing
Kondo and RKKY interactions result in a quantum phase transition (QPT)
without the interference of any other phases. Such a situation is
observed experimentally in $CeCu_{1-x}Au_x$ \cite{lohneysen,schroder}
and $YbRh_2Si_2$ \cite{gegenwart}.

Two different scenarios have been proposed for the heavy Fermion
QPT. One is the Hertz-Millis-Moriya theory \cite{millis}, which is
applicable when the energy scale of the Kondo screening is much higher
than that of the magnetic ordering near the quantum critical point
(QCP). As a result, the local moments are fully screened in the
quantum critical regime. However, by comparing the predicted critical
exponents with the experiments, the theory is found to be unapplicable
in many cases \cite{stewart,coleman,schroder,gegenwart}. In the second
scenario, one expects the magnetic ordering ($T_N$) and lattice Kondo
screening ($T_0$) energies vanish simultaneously at the QCP
\cite{coleman,schroder} (see also \cite{senthil}). Since the critical
spin fluctuations at $k=2k_F$ and $k=0$ are associated with the $T_N$
and vanishing $T_0$, respectively, it is interesting to know what kind
of critical magnetic mode may develop in the neighborhood of the
QCP. We will show later that near the QCP the two critical modes
strongly interact with each other and the critical spin fluctuation
becomes local in space (see also \cite{coleman}). In this scenario,
the local moments survive at the QCP.

It is important that the heavy Fermion physics as we understand be
obtained from a microscopic model. To this end the main theoretical
difficulty lies in treating on equal footing the Kondo and RKKY
interactions. Many bosonic mean field theories, like the
Hertz-Millis-Moriya \cite{millis} and slave-boson \cite{slaveb}
theories, fail because they rely on the order parameter of either the
magnetic or Kondo phase and miss the properties of the other. On the
other hand, a fermionic mean field theory, like the dynamical mean
field theory (DMFT) \cite{georges}, would allow the possible orders to
develop and compete and is more desirable.

The DMFT extends the Weiss mean field theory to describe fermions. The
single impurity DMFT was applied to both the Kondo and magnetically
ordered phases in heavy Fermions \cite{jarrell} (see also
Ref.\cite{varma}). Besides the over estimation of the Neel temperature
due to the lack of the magnetic fluctuation, this approach can not
capture properly the renormalization of the RKKY interaction. A
partial solution to this problem is to add an RKKY interaction to the
model and allow the renormalization of the RKKY by extending the Weiss
approximation to the interaction \cite{si}. The resulted formalism,
the so-called extended DMFT (EDMFT), is able to describe qualitatively
the heavy Fermions \cite{ping1} in both the Kondo and
antiferromagnetic (AFM) phases. However, it predicts a first order
phase transition due to the local mean field treatment of the RKKY
interaction \cite{ping1,ping2}.

A parallel path in studying the heavy Fermions is through the two
impurity problem. This model contains the dynamics of both the Kondo
and RKKY interactions and is solvable \cite{jones,affleck}. It was
shown, at the particle-hole symmetry, there was a non-Fermi liquid
fixed point separating the Kondo and magnetic phases. However, it is
difficult to extend the properties of a multi-impurity model to a
lattice of impurities.

In this letter, we combine the two impurity model with DMFT so that a
lattice of impurities can be described. This overcomes the
difficulties of both the single impurity DMFT and multi-impurity
approaches. It handles the Kondo and RKKY interactions in a more
balanced way.

We consider the periodic Anderson model in 3D

\[
  H=\sum_{\vec{k},\sigma} (\epsilon_{\vec{k}}-\mu)
  c^{\dagger}_{\vec{k},\sigma} c_{\vec{k},\sigma}
  + (E_f-\mu) \sum_{j,\sigma} n^f_{j,\sigma}
\]
\begin{equation}
\label{eq-01}
  + U \sum_{j} \left(n^f_{j,\uparrow}-\frac{1}{2}\right)
  \left(n^f_{j,\downarrow}-\frac{1}{2}\right)
  +V \sum_{j,\sigma} \left(f^{\dagger}_{j,\sigma} c_{j,\sigma}
  +c^{\dagger}_{j,\sigma} f_{j,\sigma} \right)
\end{equation}

\noindent with $\epsilon_{\vec{k}}=-\frac{1}{3} \sum_{i=1}^{3} \cos
k_i$. We divide the lattice into two interpenetrating sublattices, A
and B. The unit cell is then doubled. Applying the cavity method
\cite{georges}, we obtain an effective local action:

\[
  S^{0}=-\int_{0}^{\beta} d\tau \int_{0}^{\beta} d\tau'
  \sum_{\sigma}\sum_{X,Y=A,B}
\]
\[
  f^{\dagger}_{X,0\sigma}(\tau)
  \left[{\cal G}^{f}_{0\sigma}\right]^{-1}_{XY}
  (\tau-\tau')f_{Y,0\sigma}(\tau')
\]
\begin{equation}
\label{eq-02}
  +U \int_{0}^{\beta} d\tau \sum_{X=A,B} \left(n^f_{X,0,\uparrow}
  -\frac{1}{2}\right)
  \left(n^f_{X,0,\downarrow}-\frac{1}{2}\right)
\end{equation}

\noindent The Weiss function ${\cal G}^{f}_{0\sigma}$ is determined
self-consistently as follows. First, we use quantum Monte Carlo method
(QMC) \cite{hirsch} and obtain the impurity Green's function. Then,
from the Dyson equation for the impurities, we get the impurity
self-energy $[\Sigma^{imp}_{\sigma}]_{XY}$. The lattice self-energy is
constructed to be:

\[
  \Sigma^{ff}_{\sigma}(\vec{k},ip_n)=
\]
\begin{equation}
\label{eq-03}
  \left( \begin{array}{cc}
  \Sigma^{imp}_{AA,\sigma}(ip_n) & 
  -2d\epsilon_{\vec{k}} f(\vec{k}) \Sigma^{imp}_{AB,\sigma}(ip_n) \\
  -2d\epsilon_{\vec{k}} f^*(\vec{k}) \Sigma^{imp}_{AB,\sigma}(ip_n) &
  \Sigma^{imp}_{BB,\sigma}(ip_n)
  \end{array} \right)
\end{equation}

\noindent from which we obtain the local Green's function:

\[
  G_{loc,\sigma}^{ff}(ip_n) = \sum_{\vec{k}} \Bigglb[
  \left( \begin{array}{cc}
  ip_n+\mu-E_f & 0 \\
  0 & ip_n+\mu-E_f \\
  \end{array} \right)
\]
\begin{equation}
\label{eq-04}
  -\frac{V^2}{(ip_n+\mu)^2-\epsilon^2_{\vec{k}}}
  \left( \begin{array}{cc}
  ip_n+\mu & \epsilon_{\vec{k}} f(\vec{k}) \\
  \epsilon_{\vec{k}} f^*(\vec{k}) & ip_n+\mu \\
  \end{array} \right)
  -\Sigma^{ff}_{\sigma}(\vec{k},ip_n) \Biggrb]^{-1},
\end{equation}

\noindent with $f(\vec{k})=\exp(ik_x)$. By identifying the local
(within a unit cell) Green's function on the lattice with that of the
impurity model, we form a self-consistent loop
\cite{georges,cdmft}. While solving the impurity model we can measure
the z-direction spin susceptibility $\chi_{XY}(\tau)\stackrel{\rm
def}{=}\langle {\rm T_{\tau}}\;S^f_Z(X,\tau) S^f_Z(Y,0)\rangle $, with
$S^f_Z(X)=n^f_{X,\uparrow}-n^f_{X,\downarrow}$. The lattice
susceptibility is obtained in the same way as the lattice self-energy
given in Eq.(\ref{eq-03}).

\vspace*{0.5cm}

\begin{figure}[h]
\includegraphics[width=7.5cm]{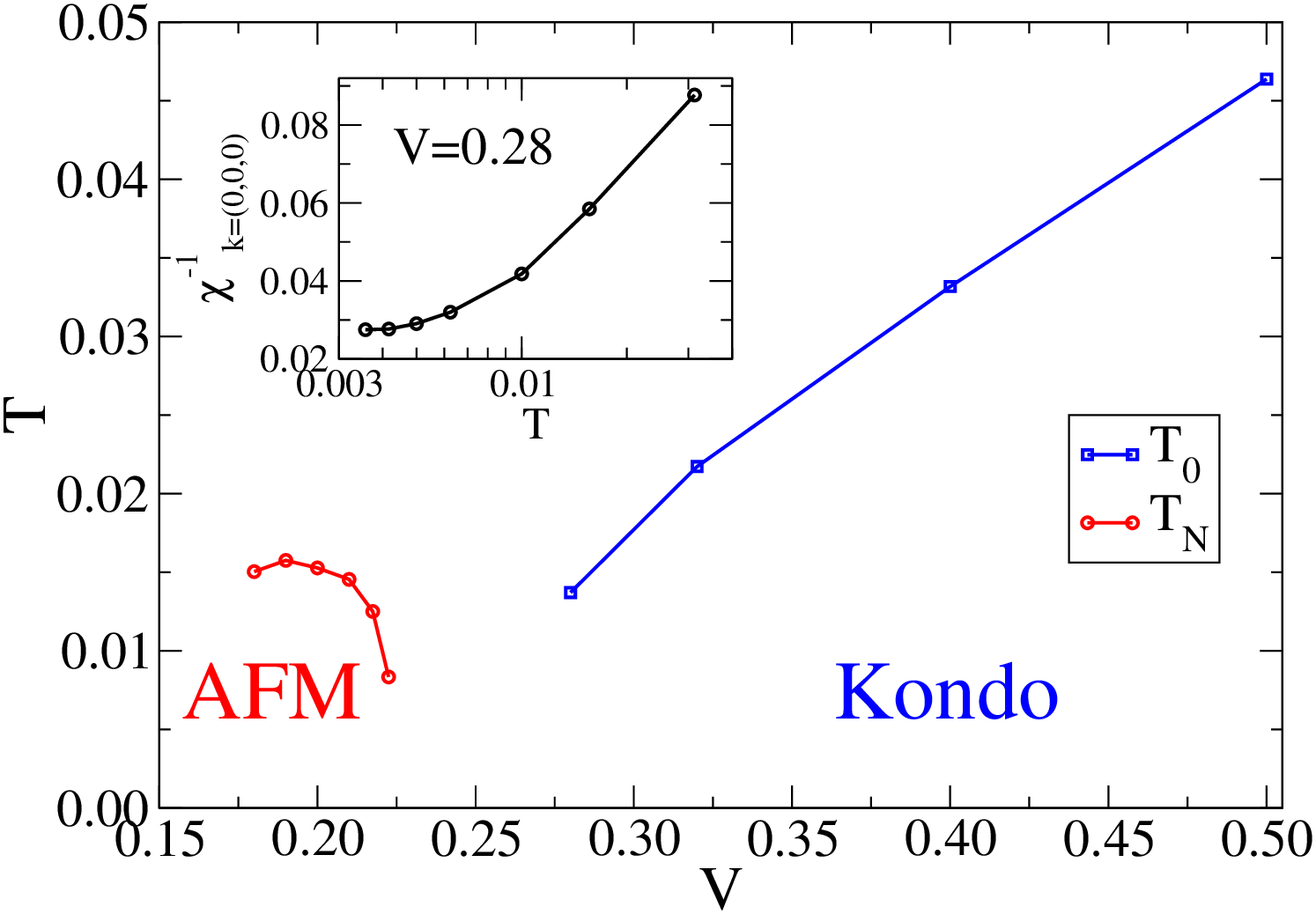}
\caption{The calculated phase diagram. The two lines, $T_N(V)$ and
$T_0(V)$, do not cross at the lowest temperatures we reached. The
inset shows the low temperature saturation of the static homogeneous
spin susceptibility at $V=0.28$.}
\label{fig-phase}
\end{figure}

We study the phase diagram of temperature v.s. $V$ at fixed $U=1.2$
and $E_f=-0.15$. To avoid crossing the band gap, we change the
chemical potential $\mu$ along with $V$ so that the free ($U=0$)
particle density per site at $T=0$ is fixed at
$N_{tot}^{Free}=2.5423$. The resulting physical density changes
slightly as $V$ increases and is always greater than and close to the
half-filling \cite{ping3}. We study the AFM and paramagetic (PM)
phases and the transition between them. In solving the two impurity
problem we use the QMC. We always use $U\Delta\tau \lesssim 1$ where
$\Delta\tau=\beta/L$ and $L$ is the number of time slices in QMC. In
each DMFT iteration, we perform QMC sweeps $\sim 10^5$. Away from the
phase transitions around 10 DMFT iterations are usually enough to
converge the results. Near the phase transition a lot more are needed
due to the critical slowing down.

Fig.\ref{fig-phase} is the phase diagram we obtained. Two technical
remarks are in place. First, to exam if the AFM to PM transition is
continuous, we checked the inverse static spin susceptibilities at
$\vec{k}=(\pi,\pi,\pi)$ which becomes very critical at the
corresponding transition values of $V$ \cite{ping3}. Second, the
crossover temperature $T_0$ is obtained using the saturated static
homogeneous spin susceptibility $\chi(\vec{k}=0,i0) \rightarrow C/T_0$
at low temperatures. [We used Curie constant $C=1/2$ which is obtained
in the high temperature limit.] An example of the saturation behavior
is shown in the inset of Fig.\ref{fig-phase}. Note that $T_0$ being
small means the FM spin fluctuation becomes very strong.

\vspace*{.6cm}

\begin{figure}[h]
\includegraphics[width=7.5cm]{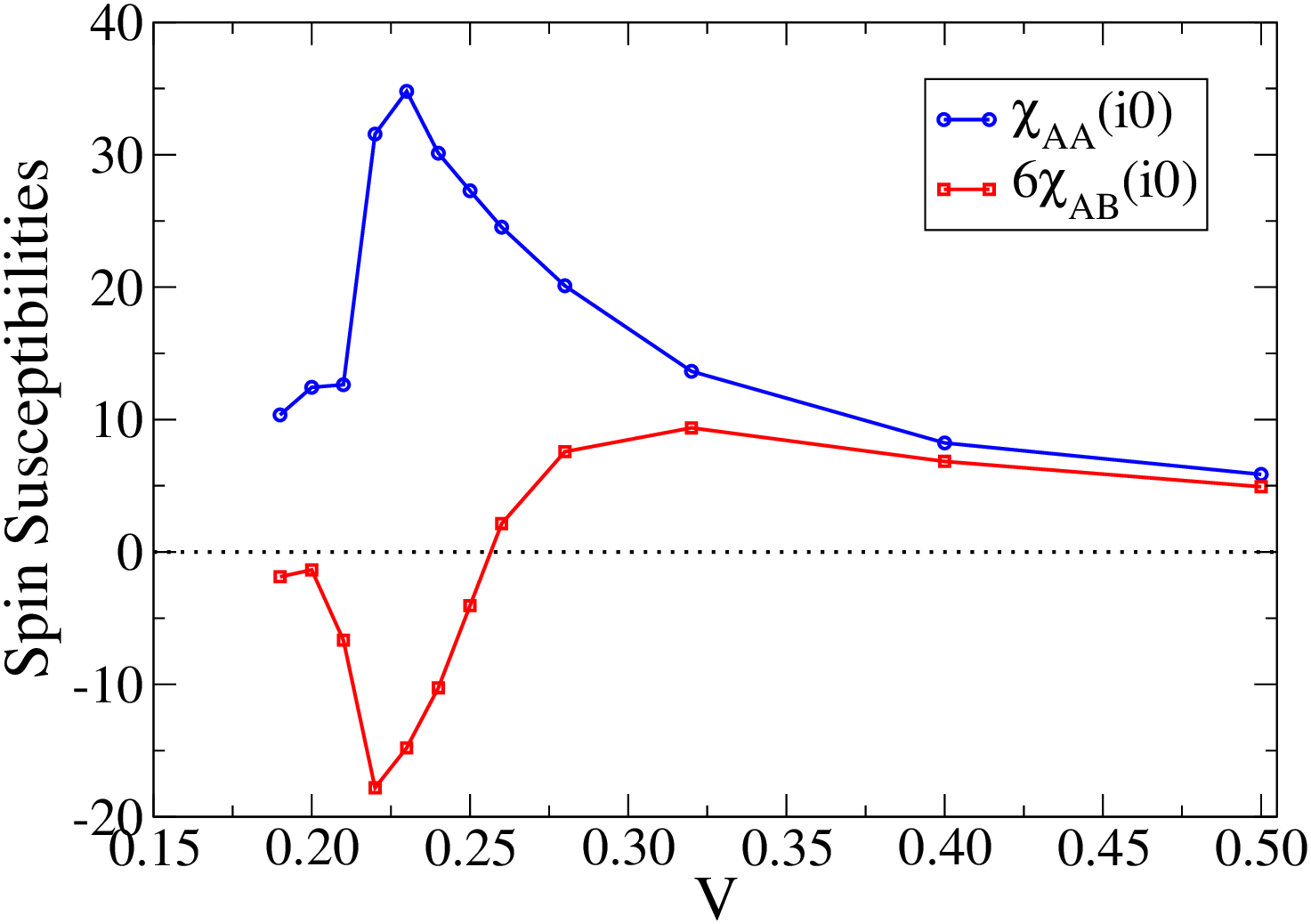}
\caption{The local, $\chi_{AA}(i0)$, and nearest neighbor,
$\chi_{AB}(i0)$, spin susceptibilities at $\beta=120$ v.s.
$V$. $\chi_{AA}(i0)$ is alway positive and shows a peak in the
crossover region. $\chi_{AB}(i0)$ changes from negative in the AFM
phase to positive in the Kondo phase. We multipled $\chi_{AB}$ by the
coordination number which reflects its contribution to the lattice
spin susceptibility.}
\label{fig-chi12}
\end{figure}

To study the critical magnetic fluctuation around the QPT, we plot in
Fig.\ref{fig-chi12} $\chi_{AA}(i0)$ and $\chi_{AB}(i0)$. From the
result we see that $\chi_{AB}(i0)$ changes sign when $V$
increases. This sign change is actually a special manifestation of a
more general evolution of the RKKY correlation from being AFM to FM as
hybridization is increased, due to the conduction electrons mediating
the RKKY change from those near the Fermi surface to around the band
bottom. Meanwhile, $\chi_{AA}(i0)$ is always positive and becomes very
strong in the crossover regime. This scenario should extend to $T=0$
and there would be a point at which $\chi_{AB}(i0)=0$ and
$\chi_{AA}(i0)$ becomes critical. This point would be the heavy
Fermion QCP. Note that the static spin response being local does not
mean a local electron self-energy. Actually, the non-local self-energy
becomes stronger at lower temperatures \cite{ping3}.

\vspace*{0.6cm}

\begin{figure}[h]
\includegraphics[width=7.5cm]{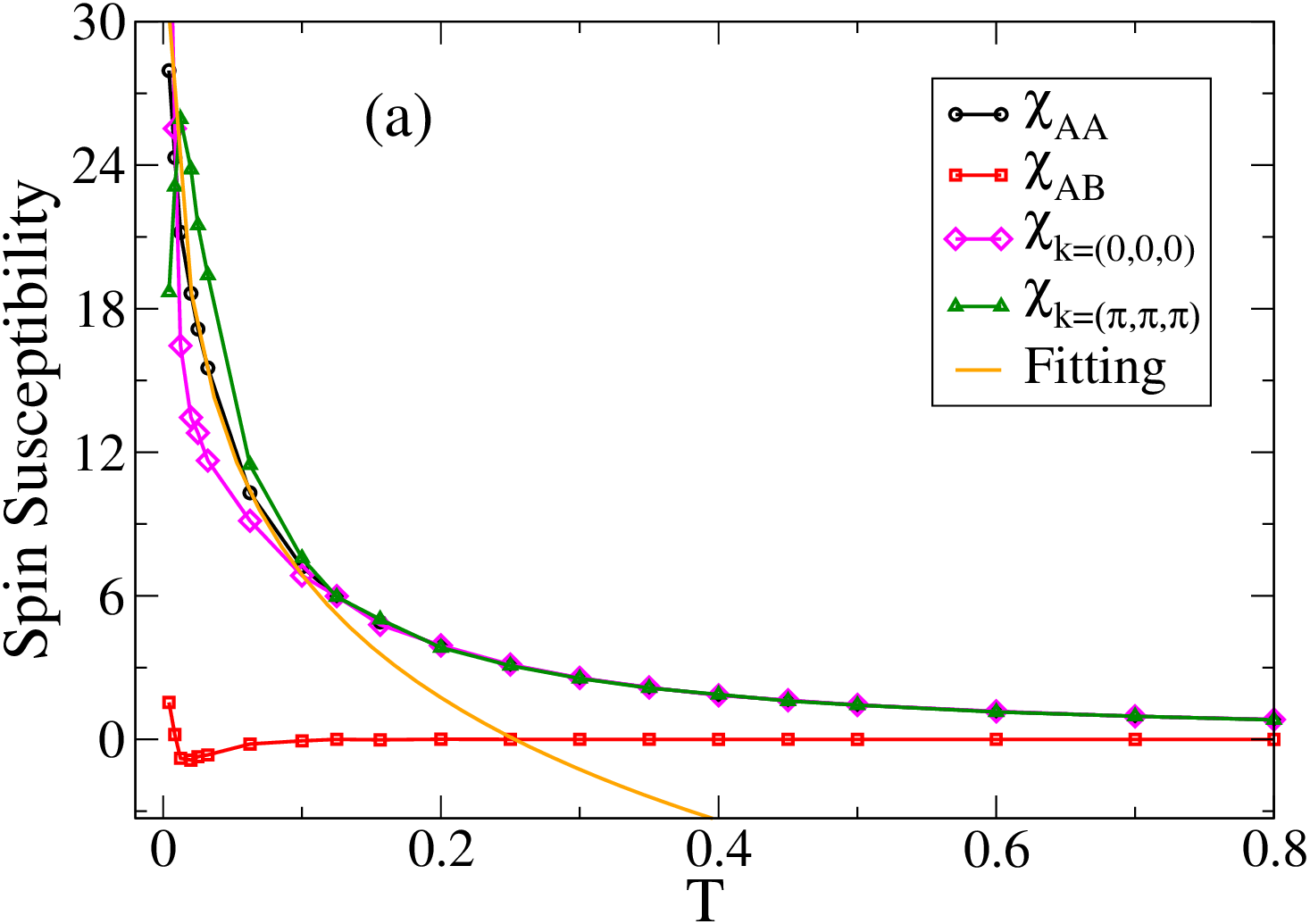}

\vspace*{0.9cm}

\includegraphics[width=7.5cm]{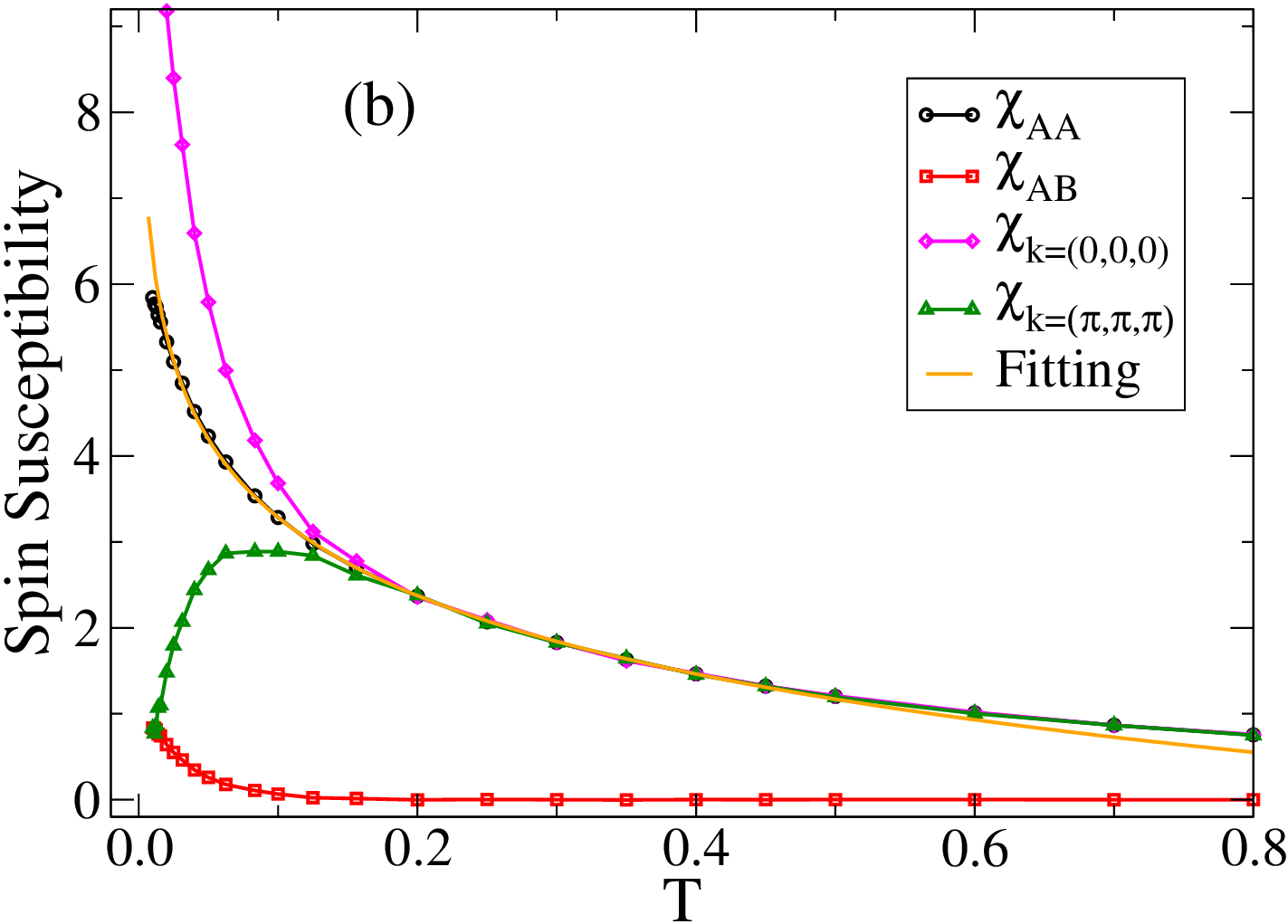}
\caption{The static spin susceptibilities as functions of the
temperature at (a) $V=0.26$ and (b) $V=0.5$. The fittings are given by
$\chi(T)=7.399 \ln(0.253/T)$ in (a) and $\chi(T)=1.314 \ln(1.218/T)$
in (b). Note that according to Fig.\ref{fig-phase} $V=0.26$ is close
to the QCP on the Kondo side.}
\label{fig-chi}
\end{figure}

\begin{figure}[h]
\includegraphics[width=7.5cm]{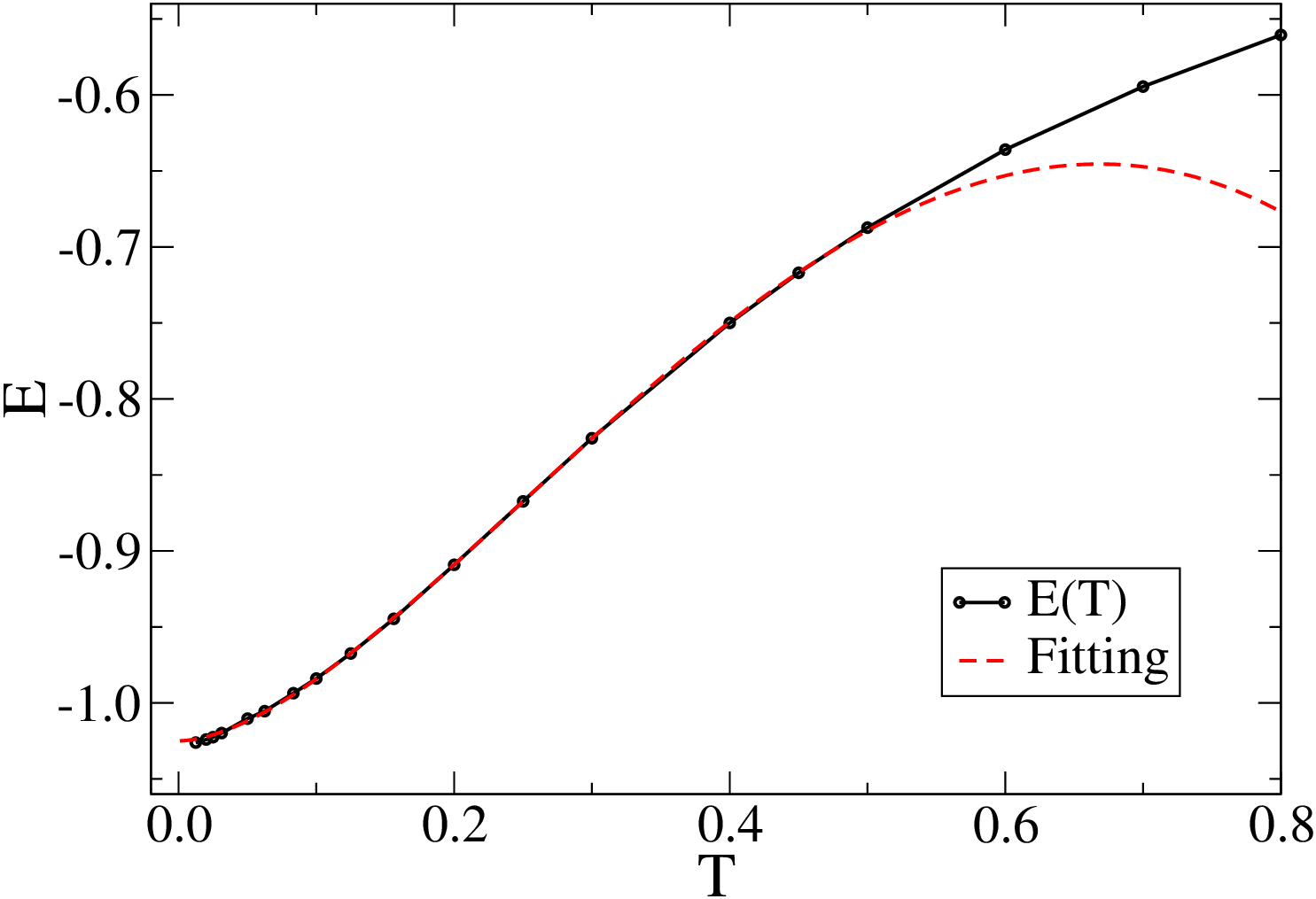}
\caption{The total energy at $V=0.26$. We fitted
$E(T)=-1.025+1.700T^2\ln(1.102/T)$. The corresponding specific heat
coefficient is $\gamma(T)=3.400\ln(0.6682/T)$.}
\label{fig-energy}
\end{figure}

An important question is how the Kondo screening, which is local in
space, becomes coherent in heavy Fermions. To this end, we study the
evolution of the spatial correlation in the spin responses in
Fig. \ref{fig-chi}. It shows that spin fluctuations are quite local in
space down to $T\sim 0.1$ for $V=0.26$ and $T\sim 0.2$ for
$V=0.50$. At lower temperatures, the FM spin susceptibility becomes
dominant. This is similar to that observed experimentally in
$YbRh_2Si_2$ \cite{ishida}. Two remarks are in place. (1) the
non-locality in $\chi$ develops at lower temperature for $V=0.26$ than
that for $V=0.50$. This reflects the local nature of the spin
fluctuation in the quantum critical region which extends to lower
temperature as the QCP is approached and is consistent with
Fig.\ref{fig-chi12}. (2) The logarithmic temperature dependence of
$\chi$ as shown in Fig. \ref{fig-chi} is similar to those observed in
many heavy Fermion compounds \cite{stewart}. A logarithmic temperature
dependence is also found in the total energy shown in
Fig. \ref{fig-energy}.

\vspace*{0.6cm}

To conclude, using a two impurity DMFT, we studied the periodic
Anderson model on cubic lattice at finite temperatures. We obtained
the phase diagram which is consistent with the picture that both the
Neel and Kondo temperatures vanish at the QCP. As the QCP was
approached from the Kondo side, we found strong ferromagnetic spin
fluctuations. From the sign change of the static nearest-neighbor spin
susceptibility, we conjectured that the critical static spin
fluctuation become local at the heavy Fermion QCP. We explored the
crossover region and observed logarithmic temperature dependences in
the specific heat coefficient and spin susceptibility.

Our results presented in this letter implies that a two impurity
Anderson model combined with DMFT might serve as a minimal model in
describing most of the thermodynamics of the heavy Fermions. However,
we should also keep in mind that there could exist other phases, like
the superconducting phase \cite{millis_sc}, whose existence may
require mechanisms which are more spatially extended than those
describable in a two impurity model.

Since the 4f or 5f orbital contributes to the local moment physics in
most heavy Fermion compounds, there are inevitablely many physical
properties over simplified by the periodic Anderson model, like the
orbital degeneracy and the subsequent crystalline field
splitting. These are further complicated by the spin-orbital coupling,
lattice frustration, disorder, hybridization with other bands,
etc. Actually all these contribute to a much richer physics observed
in experiments \cite{stewart} than what we have obtained. To this end,
our current work can be considered as a useful guide to distinguish
the ``universal'' heavy Fermion features from those specific to
individual materials.

The authors would like to acknowledge helpful discussions with
E. Abrahams, P. Coleman, A. Georges, O. Parcollet, C. Pepin,
A. Schiller, and Q. Si. This research was supported by NSF under
No. DMR-0096462, the Center for Materials Theory at Rutgers
University, and (P.S.) an I2CAM travel award.

\end{document}